\documentclass[useAMS,usenatbib]{mn2e}
\usepackage{graphicx,natbib}
\usepackage{psfrag}
\usepackage{amssymb}
\usepackage{float}
\usepackage[dvips]{color}
\usepackage{tikz}

\def\aV{\mbox{$\rm A_V$}}

\def\ms{\mbox{$M_\odot$}}

\title[High-latitude embedded clusters]{Discovery of two embedded clusters with WISE in the high Galactic latitude cloud HRK 81.4-77.8}

\author[D. Camargo, E. Bica, C. Bonatto, and G. Salerno]{D. Camargo$^{1,2}$,  E. Bica$^1$, C. Bonatto$^1$, and G. Salerno$^1$\\
$^1$ Departamento de Astronomia, Universidade Federal do Rio Grande do Sul, 
Av. Bento Gon\c{c}alves 9500\\
Porto Alegre 91501-970, RS, Brazil\\
$^2$ Col{\'e}gio Militar de Porto Alegre, Minist{\'e}rio da Defesa - Ex{\'e}rcito Brasileiro, 
Av. Jos{\'e} Bonif{\'a}cio 363\\
Porto Alegre 90040-130, RS, Brazil}

\begin{document}

\pagerange{\pageref{firstpage}--\pageref{lastpage}}

\maketitle

\label{firstpage}

\begin{abstract}

Molecular clouds at very high latitude ($b>60^{\circ}$) away from the Galactic plane are rare and in general are expected to be non-star-forming. However, we report the discovery of two embedded clusters (Camargo 438 and Camargo 439) within the high-latitude  molecular cloud HRK 81.4-77.8 using WISE.  
Camargo 439 with Galactic coordinates $\ell=81.11^{\circ}$ and $b=-77.84^{\circ}$ is an $\sim2$ Myr embedded cluster (EC) located at a distance from the Sun of $d_{\odot}=5.09\pm0.47$ kpc. Adopting the distance of the Sun to the Galactic centre $R_{\odot}=7.2$ kpc we derive for Camargo 439 a Galactocentric distance of $R_{GC}=8.70\pm0.26$ kpc and a vertical distance from the plane of $-4.97\pm0.46$ kpc. Camargo 438 at $\ell=79.66^{\circ}$ and $b=-78.86^{\circ}$ presents similar values. 
The derived parameters for these two ECs put HRK 81.4-77.8 in the halo at a distance from the Galactic centre of $\sim8.7$ kpc and $\sim5.0$ kpc from the disc.
Star clusters provide the only direct means to determine the high latitude molecular cloud distances. The present study shows that the molecular cloud HRK 81.4-77.8 is currently forming stars, apparently an unprecedented event detected so far among high latitude clouds.
We carried out a preliminary orbit analysis.
It shows that this ECs are the most distant known embedded clusters from the plane and both cloud and clusters are probably falling ballistically from the halo onto the Galactic disc, or performing a flyby. 

\end{abstract}


\section{Introduction}
\label{Intro}

Intermediate and high latitude molecular clouds (HLCs) are small and low gas density structures that may be in the transition between molecular to atomic clouds \citep{Sakamoto02}. Most of them appear to be non-star-forming clouds \citep{Magnani00}. Their origins are still not well understood, but a possible explanation is that violent events as supernovae explosions within the Galactic disc may throw dust away, which during the free fall phase can merge to form molecular/dust clouds. This model is known as Galactic fountain \citep{Shapiro76, Bregman80}. \citet{Melioli08} suggest that a typical fountain powered by 100 Type II supernovae from a single OB association may eject material up to $\sim2$ kpc \citep{Quilis01, Pidopryhora07}. 
However, an extragalactic origin is also possible \citep{Oort70, Kaufmann06} with the infall gas condensing to form clouds, which fall to the Galactic disc \citep[see also,][]{Lockman08, Nichols14}.

Star formation in very high Galactic latitude molecular clouds provides the only direct means to determine their distances, but as far as we are aware
no such  an event has been detected.

\citet{Blitz84} identified 457 HLCs (including intermediate latitudes) candidates and \citet{Magnani96} constructed a catalogue of about 100 clouds. In the absence of star formation, distances to the Galactic plane and  Sun can only be estimated by means of statistical modeling. 
One observational approach is spectral
absorptions by a foreground hot star \citep[e.g.][]{Danly92}. An alternative
is photometric detection of reddening of stars along the line of sight \citep[e.g.][]{Schlafly14}. However, the lack of direct cloud
distances remains a challenge. 
The only detected star forming cases at intermediate Galactic latitudes are MBM 12 \citep{Luhman01}, which is a small association at $b=-33.8^{\circ}$, and MBM 20 \citep{McGehee08, Malinen14} at $b=-36.5^{\circ}$.

\begin{figure}
\resizebox{\hsize}{!}{\includegraphics{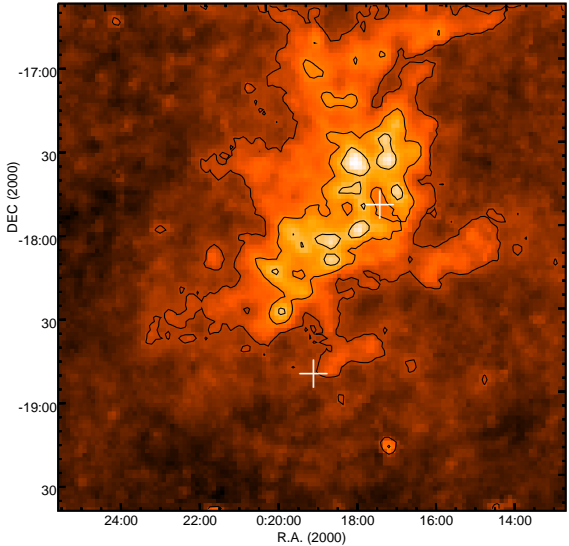}}
\put(-95.0,131.0){\makebox(0.0,0.0)[5]{\fontsize{18}{18}\selectfont \color{white} +}}
\put(-120.2,67.0){\makebox(0.0,0.0)[5]{\fontsize{18}{18}\selectfont \color{white} +}}
\caption[]{IRAS $100{\mu}m$ ($2.5^{\circ}\times2.5^{\circ}$) image showing in detail the structure of HRK $81.4-77.8$. The plus signs indicate the position of C 439 (\textit{top}) and C 438 (\textit{bottom}).}
\label{f1}
\end{figure}

High velocity clouds (HVCs) were discovered in 1963 \citep{Muller63}.
Models of HVCs also  naturally predict lower radial velocity halo clouds (LVHC). Similarly to HVCs,   expectations are that IRAS fluxes tend to be lower as compared to HI column densities in LVHCs. \citet{Peek09}  found LVHCs based on such low dust to gas ratios.

In this paper we communicate the discovery of two embedded clusters in the direction of the low velocity high latitude molecular/HI cloud HRK 81.4 -77.8 \citep{Heiles88},
also known as G $81.4-77.8$ or Gal $081.40-77.80$ - see e.g. Simbad\footnote{http://simbad.u-strasbg.fr/simbad/sim-fcoo}. \citet{Heiles88} provided a radial velocity of $v_{LSR}=-8$ km/s. The equatorial position is $\alpha(2000)=0^{h}17^{m}32^{s}$ and $\delta(2000)=-17^{d}43'18''$.
HRK 81.4-77.8 is an isolated molecular cloud located in Cetus (Heiles et al. 2009). No distances are available for HRK 81.4-77.8. We detected the embedded clusters (ECs) with WISE, and by means of stellar  photometry with the 2MASS and WISE catalogues, we computed their Colour-Magnitude Diagrams (CMDs) and radial density profile (RDP).

The two new clusters are Camargo 438 and Camargo 439, hereafter C 438 and C 439, respectively. The cluster designation  and numbering follow the recent catalogue of young clusters that we found in WISE \citep{Camargo15}.

The systematic detection of the PMS stellar content in embedded clusters have become a major achievement by our group \citep[e.g.][]{Bica08, Bica11, Bonatto10, Bonatto11, Camargo09, Camargo10, Camargo11, Camargo12, Camargo13, Camargo15}. What characterizes our analysis is decontamination of field stars.
Recently, we dedicated attention  to  embedded clusters essentially seen in WISE images \citep{Camargo15}. We discovered 437 such clusters in the Galactic disk.
We now turn our attention to high latitude clouds ($b>60^{\circ}$), knowing the capacity of our approach, and the lack as yet of direct distances and star formation for halo clouds.

In Sect.~\ref{sect2} we communicate the 2 new clusters and carry out CMD and RDP analyses. In Sect.~\ref{sect3} compute and discuss possible cloud orbits and in Sect.~\ref{sect4} we provide concluding remarks.

\begin{figure*}
\begin{center}
\begin{minipage}[b]{0.78\linewidth}
\begin{minipage}[b]{0.495\linewidth}
\includegraphics[width=\textwidth]{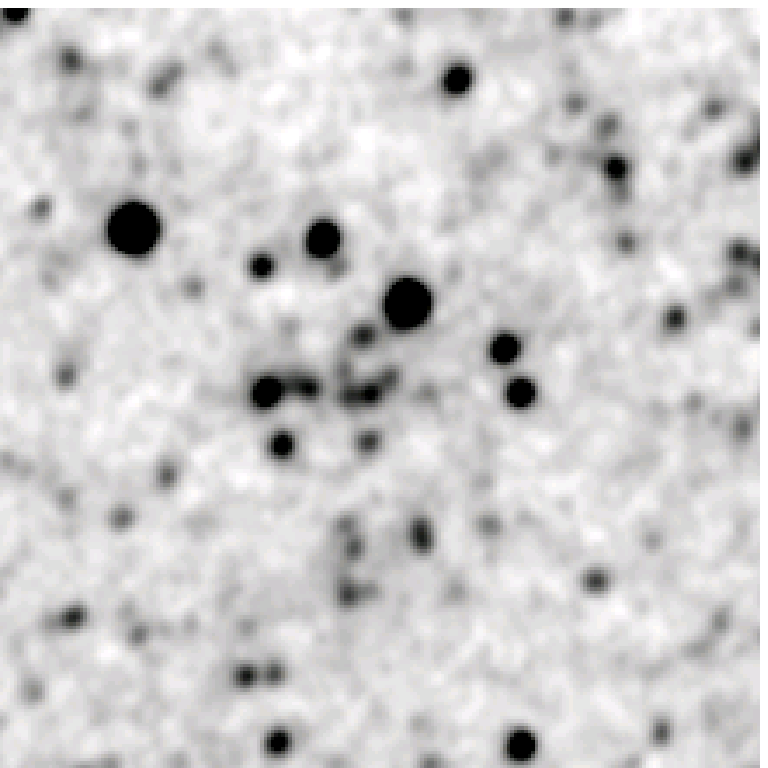}
\end{minipage}\hfill
\vspace{0.02cm}
\begin{minipage}[b]{0.495\linewidth}
\includegraphics[width=\textwidth]{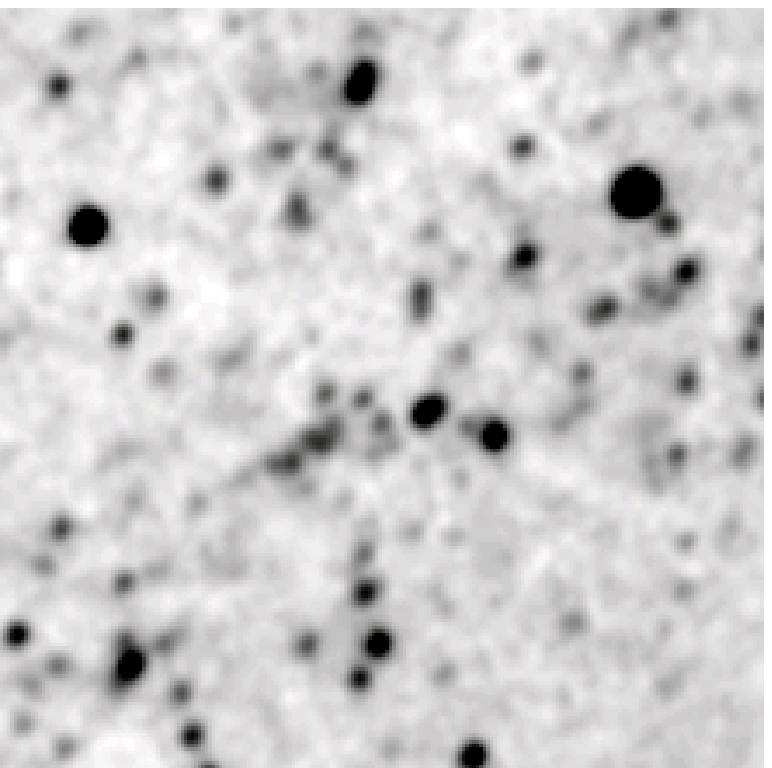}
\end{minipage}\hfill
\end{minipage}\hfill
\caption[]{The new embedded clusters. \textit{Left}: WISE W1 ($5'\times5'$) image centred on the C 439 coordinates. \textit{Right}: the same for C 438.}
\end{center}
\label{f2}
\end{figure*}

\section{Discovery of two young star clusters}
\label{sect2}
In this work we communicate the discovery of two ECs in the high-latitude molecular cloud HRK 81.4-77.8. The existence of ECs at large distances from the Galactic plane is of considerable importance because it may provide information about the physical conditions and processes that are taking place in  HLCs and how such a halo component behaves in the Galaxy. These ECs may provide a direct measurement of the scale height and distance to HLCs. Following our recent catalogue \citep{Camargo15} we adopt the designations C 438 and C 439 for the newly discovered ECs.

In the Galaxy most young open clusters and embedded clusters are located within the range of 200 pc from the Galactic plane \citep[e.g.][]{Camargo13}.

\begin{figure}
\resizebox{\hsize}{!}{\includegraphics{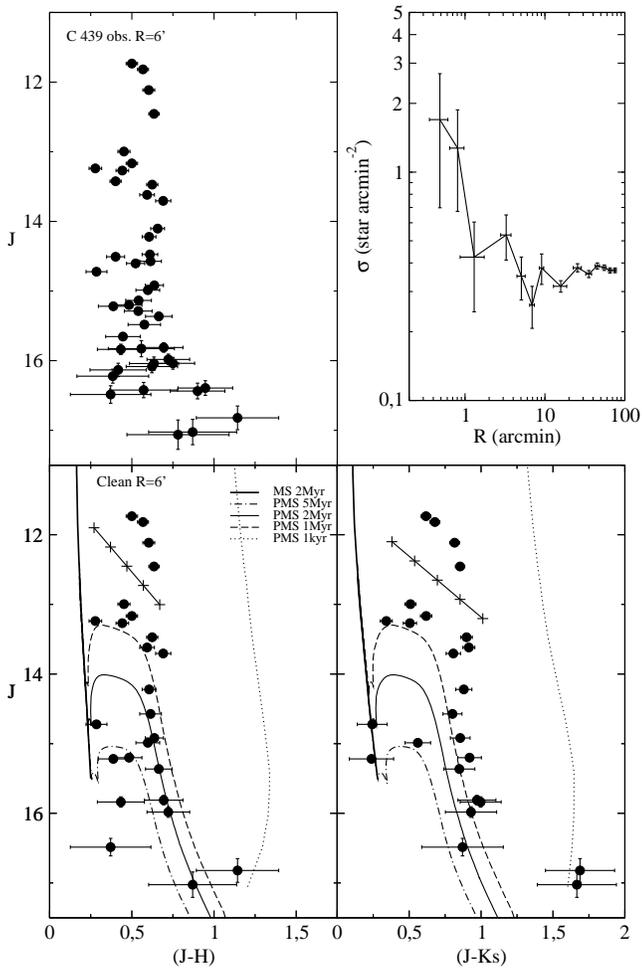}}
\caption[]{2MASS CMDs and RDP for the newly found embedded cluster C 439. Top panels: observed CMD $J\times(J-H)$ (\textit{left}) and RDP (\textit{right}). Bottom panels: field-star decontaminated CMDs fitted with Padova isochrones. We also show the reddening vector for $A_v=0$ to $5$.}
\label{f3}
\end{figure}

Fig.~\ref{f1} shows in detail the IRAS $100{\mu}m$  image of the molecular/HI cloud HRK 81.4-77.8. The cluster C 439 is located within the central region of a ring-like structure of dense gas clumps. Cloud structures like that are common around ECs and are in general related to feedback of massive stars. 
C 438 is located at the southern border of the molecular cloud.
Given the relative isolation of HRK 81.4-77.8, halo stars are probably not the driving mechanism responsible for its star formation. Besides, the present star formation is most likely
the first, since the most massive stars have not yet reached the main sequence (see Fig.~\ref{f3}). 

The WISE bands W1 ($3.4{\mu}m$) and W2 ($4.6{\mu}m$) are more sensitive to the stellar component while W3 ($12{\mu}m$) and W4 ($22{\mu}m$) show rather dust emission.
In Fig.~\ref{f2} we show WISE W1 images of the newly found ECs. We show in Table~\ref{tab1} the equatorial positions of C 438 and C439. The Galactic coordinates of C 439 are $\ell=81.11^{\circ}$ and $b=-77.84^{\circ}$ and for C 438 are $\ell=79.66^{\circ}$ and $b=-78.86^{\circ}$. The cloud  is present in the WISE W3 and W4 bands. 
Star formation in the cloud is expected to be related to cold and warm dust emission, since massive clusters may produce more massive (hot) stars that, in turn, heat the surrounding dust to higher temperatures than those expected in less massive clusters.

\begin{figure}
\resizebox{\hsize}{!}{\includegraphics{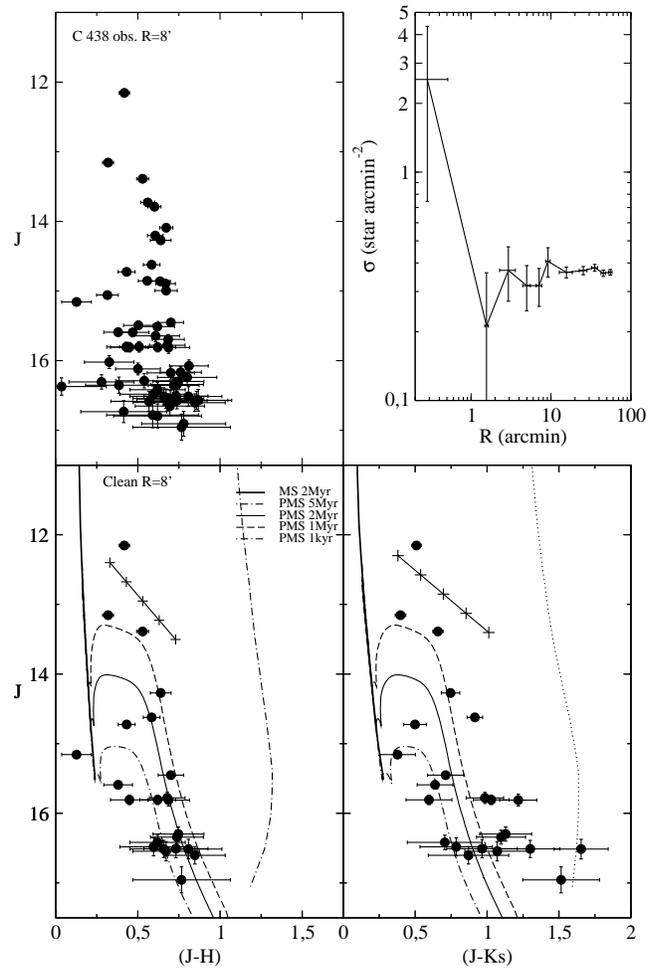}}
\caption[]{Same as Fig.~\ref{f3} for C 438.}
\label{f4}
\end{figure}

In Figs.~\ref{f3} and \ref{f4} are shown CMDs  and RDPs for both ECs following our previous studies. The CMDs are built with 2MASS photometry. The upper-left panels give CMDs extracted from a circular area centred on the coordinates of each EC. The upper-right panel of each figure presents the RDP for the respective cluster. The bottom panels give the decontaminated CMDs built applying the field star decontamination  algorithm to the raw photometry. It is described in detail in \citet{Bonatto07b, Bonatto08, Bonatto10} and \citet{Bica08} and has been used in several works \citep[e.g.][and references therein]{Camargo09, Camargo10, Camargo11, Camargo12, Camargo13, Bica11, Bonatto09, Bonatto11a}. The fundamental parameters (\textit{age, reddening, distance}) are derived by fitting PARSEC isochrones \citep{Bressan12} to the cluster sequences in the decontaminated CMD. 
The fits are made by eye allowing for differential reddening and photometric uncertainties \citep{Camargo10, Camargo11}. We apply magnitude and colour shifts to the MS+PMS isochrone set from zero distance modulus and reddening until a satisfactory solution is reached.
The parameter errors have been estimated by displacing the best-fitting isochrone in colour and magnitude to the limiting point where the fit remains acceptable. The best solution for each cluster is shown in Table~\ref{tab1}.

\begin{figure}
\resizebox{\hsize}{!}{\includegraphics{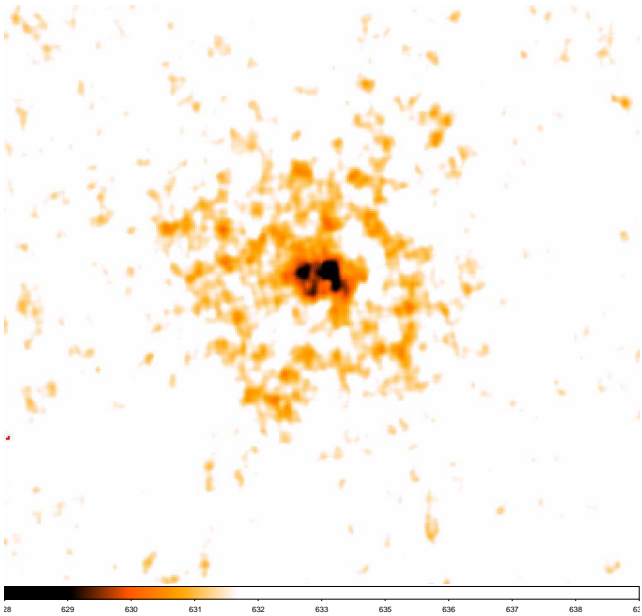}}
\caption[]{WISE W3 ($10'\times10'$) image of dust emission centred on C 438.}
\label{f5}
\end{figure}

\begin{table*}
{\footnotesize
\begin{center}
\caption{Position and derived fundamental parameters for the 2 embedded clusters.}
\renewcommand{\tabcolsep}{1.2mm}
\renewcommand{\arraystretch}{1.5}
\begin{tabular}{lrrrrrrrrrrr}
\hline
\hline
Cluster&$\alpha(2000)$&$\delta(2000)$&$\aV$&Age&$d_{\odot}$&$R_{GC}$&$x_{GC}$&$y_{GC}$&$z_{GC}$&N&M\\
&(h\,m\,s)&$(^{\circ}\,^{\prime}\,^{\prime\prime})$&(mag)&(Myr)&(kpc)&(kpc)&(kpc)&(kpc)&(kpc)&(stars)&($M_{\odot}$)\\
($1$)&($2$)&($3$)&($4$)&($5$)&($6$)&($7$)&($8$)&($9$)&($10$)&($11$)&($12$)\\
\hline
C 438 &00:19:17&-18:47:55&$0.99\pm0.03$&$2\pm1$&$5.09\pm0.70$&$8.69\pm0.40$&$-07.04\pm0.02$&$+0.97\pm0.13$&$-4.99\pm0.69$&$33$&$56$\\
C 439 &00:17:30&-17:49:18&$0.99\pm0.03$&$2\pm1$&$5.09\pm0.47$&$8.70\pm0.26$&$-07.05\pm0.02$&$+1.06\pm0.10$&$-4.97\pm0.46$&$42$&$260$\\
\hline
\end{tabular}
\begin{list}{Table Notes.}
\item Cols. 2 and 3: Central coordinates; Col. 4: $A_V$ in the cluster's central region. Col. 5: age, from 2MASS photometry. Col. 6: distance from the Sun. Col. 7: $R_{GC}$ calculated using $R_{\odot}=7.2$ kpc as the distance of the Sun to the Galactic centre. Cols. 8 - 10: Galactocentric components. Cols. 11 - 12: number of candidate cluster members and cluster mass.
\end{list}
\label{tab1}
\end{center}
}
\end{table*}

\begin{table*}
\centering
{\footnotesize
\caption{Stars with UCAC4 proper motion in the direction of C 439.}
\label{tab2}
\renewcommand{\tabcolsep}{0.7mm}
\renewcommand{\arraystretch}{1,2}
\begin{tabular}{lrrrrrrrrr}
\hline
\hline
Star designation&$\alpha(2000)$&$\delta(2000)$&$\ell$&$b$&pmRA&pmDE&other designation&membership status\\
&(h\,m\,s)&$(^{\circ}\,^{\prime}\,^{\prime\prime})$&$(^{\circ})$&$(^{\circ})$&mas/yr&mas/yr&& \\
\hline
UCAC4 361-000373&00:17:36.834&-17:48:13.42&081.2618&-77.8432&$5.2\pm1.9$&$2.1\pm2.0$&2MASS 00173683-1748135&member\\
UCAC4	361-000371&00:17:31.560&-17:48:17.09&081.1810&-77.8309&$20.6\pm3.4$&$-20.7\pm3.2$&2MASS 00173155-1748171&no member\\ 
UCAC4 361-000366&00:17:26.087&-17:49:18.00&081.0503&-77.8307&$4.8\pm3.9$&$-4.0\pm6.9$&2MASS 00172608-1749182&member\\
\hline
\end{tabular}
}
\end{table*}

The decontaminated CMDs of C 439 (Fig.~\ref{f3}) fitted by PARSEC isochrones provide an age of $2\pm1$ Myr for a distance from the Sun of $5.1\pm 0.5$ kpc. Adopting $R_{\odot}=7.2$ kpc \citep{Bica06} we derive a Galactocentric distance of $R_{GC}=8.70\pm0.26$ kpc with spatial components $x_{GC}=-7.05\pm0.02$ kpc, $y_{GC}=1.06\pm0.10$ kpc, and $z_{GC}=-4.97\pm0.46$ kpc. Nevertheless, if we adopt $R_{\odot}=8.0$ kpc we derive $R_{GC}=9.34\pm0.25$ kpc with spatial components $x_{GC}=-7.83\pm0.02$ kpc, $y_{GC}=1.06\pm0.10$ kpc, and $z_{GC}=-4.97\pm0.46$ kpc.  Usually we analyse the cluster structure by fitting a King-like profile to the cluster RDP \citep{King62}. However, the RDP of C 439 is irregular and does not follow a King's profile, which is expected for such young cluster. Nevertheless, it was possible to estimate the probable cluster radius as $R_{RDP}\sim10\,arcmin$.

The analysis of the decontaminated CMDs of C 438 (Fig.~\ref{f4}) provide an age of $2\pm1$ Myr and a distance of $d_{\odot}=5.1\pm 0.7$ kpc. We derive $R_{GC}=8.69\pm0.4$ kpc and spatial components $x_{GC}=-7.04\pm0.02$ kpc, $y_{GC}=0.97\pm0.13$ kpc, and $z_{GC}=-4.99\pm0.69$ kpc considering $R_{\odot}=7.2$ kpc. Adopting $R_{\odot}=8.0$ kpc we derive a Galactocentric distance of $R_{GC}=9.33\pm0.37$ kpc with spatial components $x_{GC}=-7.82\pm0.02$ kpc, $y_{GC}=0.97\pm0.13$ kpc, and $z_{GC}=-4.99\pm0.69$ kpc.  The RDP of C 438 is as well irregular, but shows a central peak. We estimate a cluster radius of $R_{RDP}\sim12\,arcmin$. Fig.~\ref{f5} shows dust emission in the central region of C 438 with the WISE W3 band, which confirms its embedded nature.

We estimate the cluster mass by counting stars in the decontaminated CMD (within the region $R<R_{RDP}$) of each EC. For the MS, the stellar masses are estimated from the mass-luminosity relation implied by the respective isochrone solutions, while for the PMS stars, we adopted an average mass value, as follows. Assuming that the mass distribution of the PMS stars follows \citet{Kroupa01} MF, the average PMS mass - for masses within the range $0.08\la m(\ms)\la7$ is $<m_{PMS}>\approx0.6\ms$ \citep[see][]{Bonatto10, Camargo11}. The estimated mass and probable custers members are shown in Table~\ref{tab1}. 

HRK 81.4-77.8 is located in the annular region where, according to \citet{Kalberla07} the Galactic disc gas accretion rate presents a peak ($6<R<11$ kpc). The velocity of this high latitude cloud agrees with that derived by \citet{Kaufmann06} for the infall rotating gas cloud, for which the orbital velocity decreases as a function of height from the disc.

\begin{figure}
\begin{center}
\vspace{1cm}
\begin{minipage}[b]{0.72\linewidth}
\includegraphics[width=\textwidth]{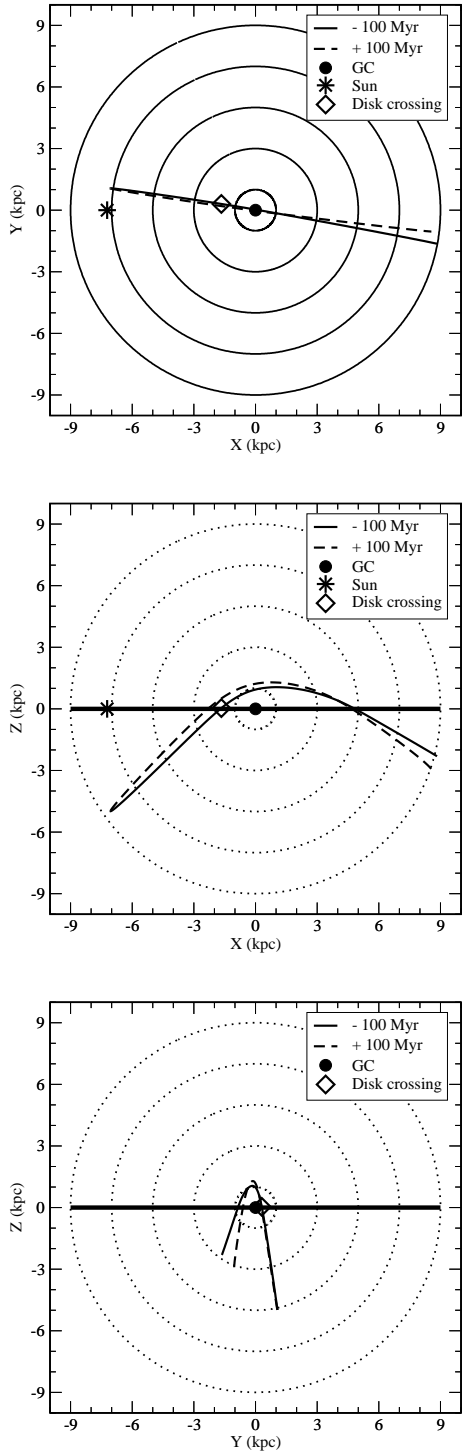}
\end{minipage}\hfill
\end{center}
\caption[]{Orbit of high latitude cloud HRK 81.4-77.8 using $V_{LSR}=-8$ km/s. Circles in the upper panel are distances in the Galactic $X-Y$ plane and in the middle and bottom panels radial distances on the vertical plane. The diamond indicates the disc crossing. The present cloud position is at the intersection of the solid and dashed lines. The Galactic disc is represented by the solid horizontal lines in the middle and bottom panels. This scenario favours a Galactic fountain model.}
\label{f6}
\end{figure}

\begin{figure}
\begin{center}
\vspace{1cm}
\begin{minipage}[b]{0.72\linewidth}
\includegraphics[width=\textwidth]{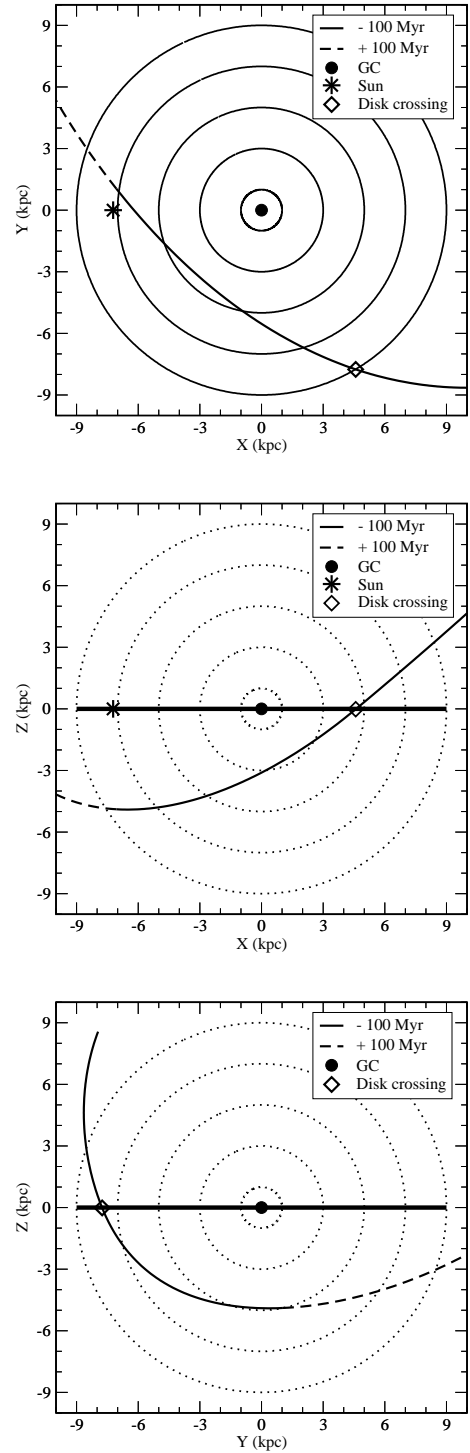}
\end{minipage}\hfill
\end{center}
\caption[]{Orbit of high latitude cloud HRK 81.4-77.8 with $V_{LSR}=-8$ km/s together with transverse velocity, considering two stars (first and third in Table~\ref{tab2}) with proper motion in the area of the cluster C 439. Symbols as in Fig.~\ref{f6}.  This scenario favours an extragalactic origin for the cloud.}
\label{f7}
\end{figure}

\section{A preliminary orbit}
\label{sect3}

The past motion of HRK 81.4-77.8  probably played an important role in the transition from atomic to molecular cloud \citep[][and references therein]{Rohser14}. \citet{Heitsch09} argue that most HI HVCs are disrupted after falling for 100 Myr and moving for $\sim10$ kpc. 
Since star-forming clouds may be successively decelerated due to fragmentation as
an infalling cloud is being disrupted, slower clouds tend to be longer-lived because
of their low mass-loss rates.
On the other hand, the disruption timescale decreases with increasing halo density, as a consequence of dragging forces. After loss of the HI content the remnants may form a warm ionized LVC in the hot coronal gas and eventually may fall in the ionized Galactic disk.

In this Sect. we compute the possible orbital motion of the molecular cloud and its two embedded clusters. For details on the Galactic potential and other procedures see \citet{Salerno09}.

We show in Fig.~\ref{f6} the cloud orbit in the Galactic potential using $V_{LSR}=-8$ km/s and null transversal velocities. This favours a Galactic Chimney model \citep{Normandeau96} for the origin of the cloud, since in this interpretation the coincidence of the cloud and the disc should have occurred $\sim48$ Myr ago. Note that this locus lies in the inner Galactic disk, where a cloud would not be expected to survive disk shocking.

In Fig.~\ref{f7} we show the cloud motion for $V_{LSR}=-8$ km/s together with  the proper motions of UCAC4 for 2 stars that have counterparts in our decontaminated CMD (Fig.~\ref{f3}). The available proper motions and other data for the stars in the area of C 439 are given in Table~\ref{tab2}. Fig.~\ref{f7} uses the proper motion of the first and third stars in Table~\ref{tab2}. Uncertainties are significant, however we tentatively compute the orbit using also this  non null components for the transverse velocity.
The orbital solution now points to an extragalactic origin for the cloud\footnote{For more definitive  results we must await GAIA (\textit{http://www.esa.int/Our\_Activities/Space\_Science/Gaia}) proper motions.}. The computation suggests a disc crossing to have occurred at $\sim46$ Myr ago, at the less dense outer disc. Such a recent shock of the cloud with the disc together with its survival would imply that the cluster formation $\sim2$ Myr ago awaited cooling processes. Employing the 2nd star in Table~\ref{tab2} the cloud would make a flyby along the Galaxy, but our simulation does not consider the deceleration by the drag of the ram pressure exerted by the halo. For the sake of emphasizing the role of uncertainties, we considered
the case of null transverse velocity, since the 1st and 3rd stars in
Table~\ref{tab2} have uncertainties of the order of the values. No UCAC4 counterpart was detected in C 438.
We conclude that HRK 81.4-77.8 is clearly a key object  for understanding the  high latitude distribution of halo clouds.  

The extragalactic clouds, especially the High Velocity ones, may have in part origin in the tidal interaction  between the  Magellanic Clouds and the Galaxy \citep{Olano04}.
The source of clouds in the  SMC/LMC would be  primarily  tidal, while in the Galactic disk the  Chimney effect may be significant. Violent star formation in the LMC and SMC will probably produce the Chimney effect.
However several hundred of massive stars in the disc are necessary to throw HRK 81.4-77.8 up to the present position, in the sense that multiple generations of star formation are required to develop a sequential supernovae event generating a continuous superwind \citep{Normandeau96, Oey05, McClure06}. Winds from OB stars may also contribute. In this way expanding superbubbles may trigger star formation generating multiple star-forming episodes renewing the fuel source needed for its own expansion and ejecting dust up even more distant \citep{Dove00, Baumgartner13}. After loss of the pressure support the dust may rain on the disc forming a Galactic fountain.

\section{Concluding remarks}
\label{sect4}

We discovered by means of WISE two embedded clusters that are located extremely far from the Galactic plane. The clusters appear to represent a star forming event in the High Latitude Low velocity molecular/HI cloud HRK 81.4-77.8.
We determined intrinsic and orbital parameters for the ensemble. As far as we are aware, this is the first detection of star formation in a high latitude molecular cloud.
The direct determination of distance for such a halo Galactic cloud is an unprecedented result.

Using WISE and 2MASS both ECs are  $\sim2$ Myr old and are located at a distance from the Sun of $\sim5.1$ kpc. Adopting $R_{\odot}=7.2$ kpc we derive for C 439 a $R_{GC}=8.7\pm0.26$ kpc and spatial components $x_{GC}=-7.05\pm0.02$ kpc, $y_{GC}=1.06\pm0.10$ kpc, and $z_{GC}=-4.97\pm0.46$ kpc. C 438 presents  $R_{GC}=8.7\pm0.4$ kpc and the spatial components $x_{GC}=-7.04\pm0.02$ kpc, $y_{GC}=0.97\pm0.13$ kpc, and $z_{GC}=-4.99\pm0.69$ kpc.

According to the derived parameters for the newly found ECs C 438 and C 439, HRK 81.4-77.8  is located at a distance from de Galactic centre of $\sim8.7$ kpc and $\sim5.0$ kpc below the disc.

In short, HRK 81.4-77.8, C 438, and C 439 may be either falling on a ballistic trajectory towards the Galactic disc, or carrying out a twisted flyby across the halo.

Existing estimates of total or main-clump masses of intermediate-latitude clouds range from $\sim30$ to $\sim220M_{\odot}$, e.g. MBM 20 (LDN 1642), MBM 41-44, and MBM 55 \citep{McGehee08, Malinen14}. Thus, our mass estimates for the embedded clusters are comparable. Considering the molecular gas to star conversion efficiency of $\sim30\%$ \citep[e.g.][]{Goodwin06}, our results imply that
high-latitude clouds should be, typically, more massive than the intermediate-latitude counterparts. A possible reason is the cloud fragmentation by halo dragging forces. The determined distance and probable orbital behaviour place HRK 81.4-77.8 in the halo, at least an order of magnitude higher than intermediate-latitude clouds.
Regarding C 438 and its location at the edge of the cloud, it might be similar to different clumps observed in intermediate-latitude clouds, such as MBM 41-44. Indeed, the $100{\mu}m$ map (Fig.~\ref{f1}) shows that HRK 81.4-77.8 hosts other dust clumps near the border as well.

\vspace{0.8cm}

\textit{Acknowledgements}: We thank an anonymous referee for valuable comments and suggestions. This publication makes use of data products from the Two Micron All Sky Survey, which is a joint project of the University of Massachusetts and the Infrared Processing and Analysis Centre/California Institute of Technology, funded by the National Aeronautics and Space Administration and the National Science Foundation. We acknowledge support from CNPq (Brazil).

\label{lastpage}
\end{document}